\documentclass[prb,aps,twocolumn,showpacs,floats,amssymb,amsfonts,amsmath]{revtex4}

\usepackage{graphicx}
\usepackage{bm}
\renewcommand{\vec}{\bm}

\begin{document}

\title{
Vorticity and vortex-core states
}

\author{
C. Berthod
}

\affiliation{
DPMC, Universit\'e de Gen\`eve, 24 quai Ernest-Ansermet,
1211 Gen\`eve 4, Switzerland
}


\begin{abstract}

The origin of the vortex-core states in $s$-wave and $d_{x^2-y^2}$-wave
superconductors is investigated by means of some selected numerical
experiments. By relaxing the self-consistency condition in the Bogoliubov-de
Gennes equations and tuning the order parameter in the core region, it is shown
that the suppression of the superfluid density in the core is not a necessary
condition for the core states to form. This excludes ``potential well'' types
of interpretations for the core states. The topological defect in the phase of
the order parameter, however, plays a crucial role. This observation is
explained by considering the effect of the vortex supercurrent on the
Bogoliubov quasiparticles, and illustrated by comparing conventional vortices
with multiply-quantized vortices and vortex-antivortex pairs. The core states
are also found to be extremely robust against random phase disorder.

\end{abstract}

\maketitle

\section{Introduction}

The vortices govern the electromagnetic response of type-II superconductors and
have been extensively studied, both experimentally and
theoretically.\cite{Blatter-94,Huebener-02} A vortex is formed from a core of
radius $r_c\sim\xi$ where the superfluid density is gradually suppressed, $\xi$
the superconducting coherence length, and it is surrounded by a supercurrent
which screens the magnetic field on a length of order $\lambda$, the
penetration depth. The vortices are strong inhomogeneities of the
superconducting condensate and they scatter the quasiparticles in several
different ways.\cite{Melnikov-01} In particular, the vortices can capture
Bogoliubov excitations into low-energy localized states.

The vortex-core states play an important role in the thermodynamic and
transport properties in the mixed state. For example, when vortices move in an
applied electric field, the core states interact with the lattice and are thus
responsible for the dissipation of energy. These states are also affected by
localized perturbations and they contribute to the pinning of vortices by
defects or impurities. Recently, mesoscopic superconducting disks have
attracted much attention.\cite{Chibotaru-00} In these systems a ``giant-vortex
state'' can be stabilized, where a single vortex at the center of the sample
carries the whole magnetic flux. In such a case the core states are the main
low-energy excitations, and they are expected to play a dominant role.
Furthermore, the strong dependence of the vortex-core energy spectrum on the
applied magnetic field opens interesting perspectives for
applications.\cite{Melnikov-02}

In $s$-wave superconductors the vortex-core bound states were predicted long
ago, based on approximate solutions of the microscopic Bogoliubov-de Gennes
(BdG) equations,\cite{Caroli-64,Bardeen-69} and subsequently observed in
NbSe$_2$ using scanning tunneling spectroscopy.\cite{Hess-89} The early
analytical results were confirmed by a complete numerical solution of the BdG
equations.\cite{Gygi-90} Extended quasiparticle excitations in the mixed state
were often studied within the quasiclassical
approach.\cite{Huebener-02,Rainer-96} Although this approximation is considered
inaccurate near the vortex core, it was used by Volovik to argue that the
number of branches of core states crossing the Fermi energy as a function of
angular momentum (considered as a continuous variable) is equal to the winding
number of the vortex.\cite{Volovik-93} This prediction was confirmed recently
by a detailed numerical solution of the BdG equations for multiply-quantized
vortices.\cite{Virtanen-99} The vortex-core states are usually thought of as
Andreev bound states, i.e. standing waves resulting from the multiple Andreev
reflection at the normal/superconductor boundary in the
core.\cite{Huebener-02,Rainer-96,Stone-96,Hofmann-98} This interpretation
suggests that the suppression of the superfluid density in the core is the main
reason for the formation of the core states. The vanishing of the superfluid
density in the core, however, plays no role in Volovik's argument. Instead, the
structure of the vortex-core energy spectrum in the approach of
Ref.~\onlinecite{Volovik-93} is entirely determined by the winding number of
the vortex, which measures the strength of the supercurrent circulating around
it. This result seems difficult to reconcile with the Andreev-bound state
picture. In particular, the peculiar dependence of the spectrum of core states
on vorticity\cite{Virtanen-99} can hardly be attributed to the order-parameter
suppression alone. Hence it is of interest to identify the roles played by the
supercurrent, on the one hand, and by the order-parameter suppression, on the
other hand, in the formation of the vortex-core states.

Based on numerical and analytical solutions of the microscopic BdG equations, I
show that the structure of the bound-state spectrum in $s$-wave and $d$-wave
vortices is determined by a topological constraint that the circulating
superfluid imposes to Bogoliubov quasiparticles. The suppression of the order
parameter in the core plays a minor quantitative role, slightly changing the
energy of the states with small angular momenta. This implies, in particular,
that a complete self-consistent treatment of the order-parameter profile is in
general not necessary, unless one is interested in detailed quantitative
predictions. These results suggest that the mechanism leading to quasiparticle
localization in vortices is quite different from other localization mechanisms
in condensed matter.

\section{Topology of vortices and BdG equations}

A vortex is a topologically stable defect of the superconducting order
parameter
$\Psi(\vec{r})\equiv\Delta(\vec{r})\,e^{i\chi(\vec{r})}$.\cite{Toulouse-76} It
is characterized by a winding number $\nu$, a topological invariant defined as
	\begin{equation}\label{eq:nu}
		\nu=\frac{1}{2\pi}\oint\vec{\nabla}\chi\cdot d\vec{l}.
	\end{equation}
The integral runs along a closed path around the vortex axis. $|\nu|$ counts
the number of $2\pi$ rotations of the phase $\chi(\vec{r})$ along the path.
Clearly $\nu$ is invariant under all continuous distortions of the phase.
Except near surfaces or interfaces, $\nu$ corresponds to the magnetic flux
$\Phi$ carried by the vortex according to $\Phi=\nu\Phi_0$ with
$\Phi_0=\frac{h}{2e}$ the superconducting flux quantum. Another characteristics
of the vortex is the shape of the order-parameter modulus $\Delta(\vec{r})$ in
the vicinity of the vortex axis, where it is constrained to vanish.

For an isolated axisymmetric vortex in a $s$-wave superconductor one has
$\Delta(r,\,\theta,\,z)\equiv\Delta(r)$ and $\chi(r,\,\theta,\,z)=\nu\theta$,
where $\nu$ is the winding number consistently with Eq.~(\ref{eq:nu}). The
modulus $\Delta(r)$ vanishes as $r\rightarrow0$, and approaches the constant
value $\Delta_{\infty}$ at $r>r_c$, where $r_c\sim\xi\ll\lambda$ in type-II
superconductors. The excitation spectrum of the vortex is determined by the
Bogoliubov-de Gennes (BdG) equations,
	\begin{equation}\label{eq:BdG}
		\left(\begin{matrix}\hat{h}&\Psi\\
		\Psi^{\star}&-\hat{h}^{\star}\end{matrix}\right)
		\left(\begin{matrix}u\\v\end{matrix}\right)=
		E\left(\begin{matrix}u\\v\end{matrix}\right),
	\end{equation}
where $\hat{h}=\frac{1}{2m}(\vec{p}-e\vec{A})^2-E_{\text{F}}$ and $u$ ($v$) is
the electron (hole) wave function of the excitation. In order to solve
Eq.~(\ref{eq:BdG}) one usually eliminates the phase $\chi$ of the order
parameter from the off-diagonal terms by performing the gauge transformation
$\vec{A}\rightarrow\vec{A}-\frac{\hbar}{2e}\vec{\nabla}\chi$. The order
parameter changes according to $\Psi\rightarrow\Psi e^{-i\chi}$ and thus
becomes real.\cite{Caroli-64} This transformation is achieved by writing the
wave functions as
	\begin{equation}\label{eq:uv}
		\left[\begin{matrix}u(\vec{r})\\v(\vec{r})\end{matrix}\right]=\left[
		\begin{matrix}e^{ik_zz}e^{i(\mu+\frac{\nu}{2})\theta}\psi_+(r/\xi)\cr
		e^{ik_zz}e^{i(\mu-\frac{\nu}{2})\theta}\psi_-(r/\xi)\end{matrix}\right].
	\end{equation}
The phase $\pm\frac{\nu}{2}\theta$ in Eq.~(\ref{eq:uv}) is an Aharonov-Bohm
phase, reflecting the fact that the gauge function $-\frac{1}{2}\chi$ carries a
singular magnetic field at the origin.\cite{Tesanovic-00} Furthermore, the
quantum number $\mu$ must be such that the total phase accumulated by the
electron and hole upon a $2\pi$ rotation around the origin is consistent with
the enclosed flux, i.e.
	\begin{equation}\label{eq:mu}
		\mu=n+\frac{\nu}{2},\qquad n\text{ integer}.
	\end{equation}
Inserting Eq.~(\ref{eq:uv}) into Eq.~(\ref{eq:BdG}) solves for the $\theta$ and
$z$ dependencies and leads to the radial equations for the real functions
$\psi_{\pm}$:
	\begin{subequations}\label{eq:radial}
	\begin{eqnarray}
		\label{eq:radial1}
		&&\pm L_{\pm}\psi_{\pm}(\rho)+\delta(\rho)\psi_{\mp}(\rho)=
			\varepsilon\psi_{\pm}(\rho)\\
		\label{eq:L}
		&& L_{\pm}=\frac{1}{g^2}\left[-\frac{d^2}{d\rho^2}-\frac{1}{\rho}
			\frac{d}{d\rho}+\frac{(\mu\pm\frac{\nu}{2})^2}{\rho^2}\right]-1.
	\end{eqnarray}
	\end{subequations}
Here $g=k_{\text{F}}\xi$ and I have introduced the dimensionless variables
$\rho=r/\xi$, $\varepsilon=E/E_{\text{F}}$, and
$\delta(\rho)=\Delta(\rho\,\xi)/E_{\text{F}}$. For simplicity I restricted in
Eq.~(\ref{eq:L}) to the two-dimensional case by putting $k_z=0$. The vector
potential $\vec{A}$ was also omitted since it is small in the core region
compared to the gauge field $\frac{\hbar}{2e}\vec{\nabla}\chi$ when
$\lambda\gg\xi$.\cite{Caroli-64} Indeed $A\sim\frac{\nu\Phi_0}
{4\pi\lambda^2}r$, and the ratio of $A$ to the gauge field is thus of order
$(r/\lambda)^2$.

Although the gauge transformation removes the phase of the order parameter, it
does not eliminate the supercurrent from the problem. In the radial equation,
the supercurrent shows up as a central potential containing a repulsive term
$(\nu/2\rho)^2$, as well as a term $\pm\mu\nu/\rho^2$ which is either
attractive or repulsive: if $\nu<0$ (supercurrent flowing counter-clockwise,
corresponding to a positive magnetic field along the $z$ axis) this term is
attractive (repulsive) for the electrons (holes) which move alike the
superfluid ($\mu>0$). Therefore, the supercurrent acts on the electron and hole
parts of the BdG excitations in different ways, and tends to decrease the
angular momentum of the former and to increase the angular momentum of the
latter (for $\nu<0$ and $\mu>0$). This effect of the supercurrent on the BdG
excitations is central to understand the formation of the core states (see
Sect.~\ref{sect:interpretation}). Furthermore, the strength of the supercurrent
fixes the parity of the vortex-states angular momentum in Eq.~(\ref{eq:mu}),
which is half an odd integer for odd $\nu$ and integer for even $\nu$. When
$\nu$ is odd, the flux carried by the vortex is not a multiple of the flux
carried by the quasiparticle, hence a topological frustration which translates
into a branch cut discontinuity---removed by the gauge transformation in
Eq.~(\ref{eq:uv})---in the angular wave function.

Eq.~(\ref{eq:BdG}) possess the well-known particle-hole symmetry $(u,\,v,\,E)
\leftrightarrow (v^*,\,-u^*,\,-E)$. Looking at the wave functions in
Eq.~(\ref{eq:uv}), one sees that in the radial equation this symmetry becomes
$(\mu,\,\psi_+,\,\psi_-,\,\varepsilon) \leftrightarrow
(-\mu,\,\psi_-,\,-\psi_+,\,-\varepsilon)$: the vortex-core energy spectrum is
invariant under the simultaneous inversion of angular momentum and energy.
Furthermore, it is clear from Eq.~(\ref{eq:radial}) that the spectrum is also
invariant under the simultaneous inversion of $\mu$ and $\nu$.

\section{The isolated $\bm{S}$-wave vortex}

Self-consistent solutions of the BdG equations for the order-parameter profile
and the energy spectrum of an isolated $s$-wave vortex were already reported,
both in the singly-quantized\cite{Gygi-90} and
multiply-quantized\cite{Virtanen-99} cases. The purpose of this section is to
repeat these calculations without achieving the self-consistency in
$\Delta(r)$, in order to clarify the role played by the detailed form of
$\Delta(r)$ and by the winding number $\nu$ in the formation of the core
states. Analytical solutions for the singly-quantized vortex also
exist.\cite{Caroli-64,Bardeen-69} I will discuss a new analytical solution
which is valid for all (integer) values of $\nu$, and which emphasizes the key
role of the winding number.

\subsection{Numerical results}

The BdG equations (\ref{eq:radial}) were solved numerically using the
Bessel-function expansion described in Ref.~\onlinecite{Gygi-90}. Besides its
winding number and order-parameter profile, an isolated vortex in a continuum
free-electron model is characterized by the bulk parameters $g=k_{\text{F}}\xi$
and $\delta=\Delta_{\infty}/E_{\text{F}}$. Physical values of $g$ range from
$\sim1$ in high-$T_c$ materials to $10$--$10^4$ in conventional
superconductors. Simulations were performed for $g$ between $1$ and $100$ (the
computational effort increases rapidly with increasing $g$). The parameters $g$
and $\delta$ are in principle related by the BCS relation
$\xi\approx\frac{\hbar v_{\text{F}}}{\pi\Delta_{\infty}}$, i.e.
$g\delta\approx2/\pi$. For each $g$, values of $\delta$ between $0.1$ and $10$
times the BCS value were considered. The reported conclusions apply to the
whole domain of parameters investigated. Below I present results for $g=10$ and
$\delta=2/(\pi g)$.\cite{note-R} For the order-parameter profile the following
form was used:\cite{Lasher-67}
	\begin{equation}\label{eq:profile}
		\Delta(r)=\Delta_{\infty}
		\tanh\left(\frac{r}{\sqrt{|\nu|}r_c}\right)^{|\nu|}.
	\end{equation}
This functional form with $r_c=\xi$ is in good qualitative agreement with the
self-consistent results of Ref.~\onlinecite{Virtanen-99}. By tuning the value
of $r_c$ one can study the effect of the gap profile on the vortex-core
spectrum. In particular, the limit $r_c\rightarrow0$ corresponds to a vortex
with no normal core. (In the remainder of this paper, I shall use the
expression {\em normal core\/} as a synonym for {\em suppression of $\Delta(r)$
in the vortex core\/}.) According to the Andreev-reflection picture, one may
expect that reducing the core radius will increase the energy separation
between the vortex-core energy levels, and eventually push the core states
outside the energy gap into the continuum as $r_c\rightarrow0$.

\begin{figure}[t!]
\includegraphics[width=7.5cm]{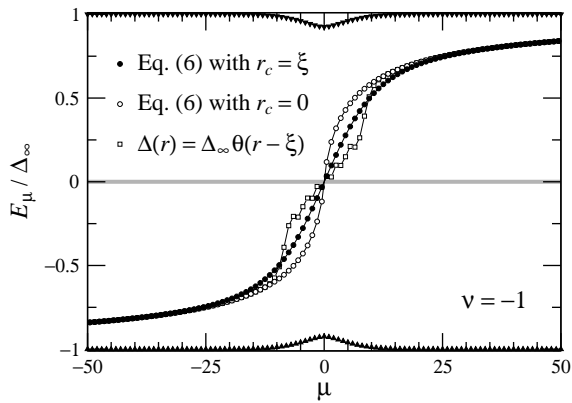}
\caption{\label{fig1}
Subgap energy spectrum of a singly-quantized $s$-wave vortex as a function of
angular momentum $\mu$ for various order-parameter profiles in the core: nearly
self-consistent profile (black circles), ``no-core'' profile (white circles),
and step-like profile (squares). The triangles show the energy spectrum of a
normal region analogous to a vortex core embedded in a uniform superconductor
(no supercurrent).}
\end{figure}

The spectra of subgap states for a singly-quantized vortex ($\nu=-1$) and for
gap profiles corresponding to $r_c=\xi$ and $r_c=0$ are shown in
Fig.~\ref{fig1}. The eigenvalues are displayed as a function of the angular
quantum number $\mu$. At energies outside the superconducting gap, the BdG
states form a continuum not shown in the figure. The spectrum obtained for a
step-like profile $\Delta(r)=\Delta_{\infty}\theta(r-\xi)$, which is often used
in analytical calculations, is also displayed for comparison. The spectra are
similar in all three cases, except for small differences at low values of
$|\mu|$. These differences have a rather simple explanation: the corresponding
eigenstates are concentrated close to the vortex axis---the maximum of the wave
function lies roughly at position $|\mu|/g$ in units of $\xi$; thus the pairing
energy of the states with $|\mu|<g$ is lowered when the order parameter gets
suppressed at $r<\xi$. It is clear from the figure, however, that the spectrum
retains its general shape even when the vortex has no normal core. In
particular, the number of branches crossing the Fermi energy is always one, in
agreement with Volovik's theorem.

We have seen that suppressing the normal core, keeping only the supercurrent,
does not change the core states qualitatively. We may now do the converse: in
order to suppress the supercurrent and keep only the core, we set the phase of
the order parameter to zero [i.e. $\nu=0$ in
Eqs.~(\ref{eq:uv})--(\ref{eq:radial})] and we use the profile
Eq.~(\ref{eq:profile}) with $\nu=1$ and $r_c=\xi$. The resulting order
parameter no longer describes a vortex, since the winding number vanishes, but
just a normal region embedded in a uniform superconductor. The resulting
spectrum (triangles in Fig.~\ref{fig1}) is qualitatively different from the
spectrum of a singly-quantized vortex. Consistently with Volovik's result, no
branch of core states cross the Fermi level, and therefore no low-energy states
exist, although multiple Andreev reflections are in principle possible in this
system. The suppression of the order parameter slightly decreases the pairing
energy of the low-$|\mu|$ electron and hole excitations of the uniform
superconductor, and gives rise to two shallow branches of states near the gap
edges.

\begin{figure}[t!]
\includegraphics[width=8.6cm]{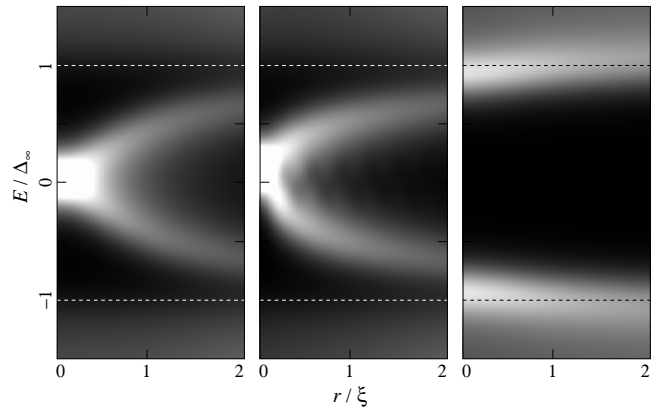}
\caption{\label{fig2}
Local density of states for the systems shown in Fig.~\ref{fig1}: conventional
singly-quantized vortex (left), vortex without normal core (center), and normal
core without supercurrent (right). A thermal broadening was used with
$k_{\text{B}}T=\Delta_{\infty}/10$.}
\end{figure}

Fig.~\ref{fig2} compares the local density of states (LDOS) of a
singly-quantized vortex with and without normal core. As can be inferred from
Fig.~\ref{fig1}, the effect of the normal core is mainly to raise the peak at
the vortex center, without changing the structure of the LDOS. The energy of
the peak is also closer to the Fermi energy when the normal core is present.
This is due to the core-induced energy change at low $\mu$. Indeed, since only
states with $\mu\pm\frac{\nu}{2}=0$ contribute to the LDOS at $\vec{r}=0$ [see
Eq.~(\ref{eq:wave})], we have
	\begin{equation}
		N(0,\,E)\propto\sum_{i}\delta\left(E-E_{\mu=-\frac{\nu}{2}}^i\right)
		\qquad(\nu\leqslant0),
	\end{equation}
where the index $i$ corresponds to the various eigenvalues at fixed $\mu$.
Fig.~\ref{fig2} also shows the LDOS of a normal core without supercurrent. In
this case there is no zero-bias peak, but two sharp peaks near
$\pm\Delta_{\infty}$ corresponding to the two branches in Fig.~\ref{fig1}.
These peaks disappear beyond a distance of order $\xi$ and the LDOS becomes
that of the uniform superconductor for $r>\xi$. In contrast, for a true vortex
the perturbation to the bulk DOS extends to distances much larger than $\xi$,
irrespective of the core radius $r_c$, due to the high-$|\mu|$ states in
Fig.~\ref{fig1}.

\begin{figure}[t!]
\includegraphics[width=7.5cm]{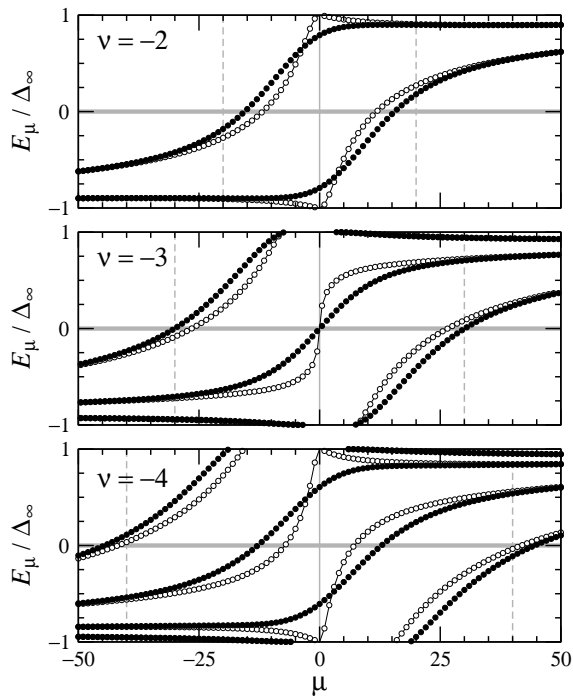}
\caption{\label{fig3}
Energy spectra for multiply-quantized $s$-wave vortices. The black circles
correspond to vortices having a normal core given by Eq.~(\ref{eq:profile}),
and the white circles correspond to vortices without normal core. The dashed
vertical lines delimit the region $|\mu|<|\nu|g$ where the levels are sensitive
to the profile of the order parameter.}
\end{figure}

The vortex-core spectra and the LDOS of multiply-quantized $s$-wave vortices
($|\nu|>1$) were calculated in Ref.~\onlinecite{Virtanen-99}, using the
self-consistent order parameter profile. The number of branches crossing
$E_{\text{F}}$ was $|\nu|$ as expected. I have repeated these calculations
using the profile Eq.~(\ref{eq:profile}) with both $r_c=\xi$ and $r_c=0$. The
results displayed in Fig.~\ref{fig3} show again that the order-parameter
profile does not change qualitatively the core states. The overall shape of the
spectra is determined by the strength of the supercurrent, which is
proportional to the winding number. For odd $\nu$, there is a branch crossing
the Fermi energy at $\mu=0$, while for even $\nu$ there is no such branch. In
addition, other branches cross zero energy at higher angular momenta. The
profile of $\Delta(r)$ affects the core states with $|\mu|<|\nu|g$. Except for
small quantitative differences, the LDOS computed with and without the normal
core are therefore very similar, as for the singly-quantized vortex.

From Figs.~\ref{fig1} and \ref{fig2} one sees that the presence of the
supercurrent is a necessary and sufficient condition to have low-energy states
(while the normal core is not), and from Fig.~\ref{fig3} one sees that the
topological frustration, which is only present for odd $\nu$, is a necessary
condition to have low-energy {\em and\/} low-$|\mu|$ states, i.e. a zero-bias
peak in the LDOS at the core center. In order to build a consistent explanation
of the origin of the vortex-core states, it is thus necessary to understand the
effect of the circulating superfluid on the BdG excitations.

\subsection{Supercurrent and pairing energy}
\label{sect:interpretation}

\begin{figure}[t!]
\includegraphics[width=7.5cm]{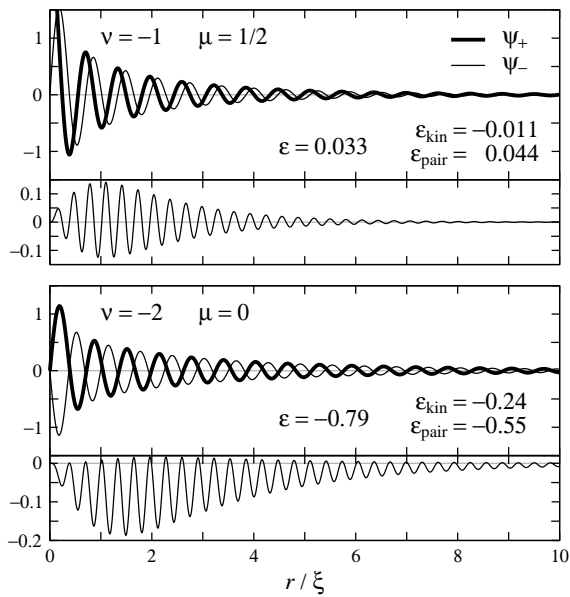}
\caption{\label{fig4}
Electron ($\psi_+$) and hole ($\psi_-$) BdG amplitudes for the lowest angular
momentum in the singly- (top) and doubly- (bottom) quantized vortex with
$r_c=\xi$. The integrand of the pairing matrix element,
$\rho\,\tanh(\rho/\sqrt{|\nu|})^{|\nu|} \psi_+(\rho)\psi_-(\rho)$, is displayed
in the bottom panels.}
\end{figure}

As already mentioned, the velocity field due to the supercurrent changes the
angular momenta of the electron and hole parts of the BdG excitations in
opposite ways. As a result, a phase difference of $\nu\frac{\pi}{2}$ is induced
between the {\em radial\/} wave functions of the electron and hole. This phase
difference is at the origin of the strong dependence of the vortex-core spectra
on $\nu$. One possible way to solve the BdG equations is to treat the modulus
of the order parameter perturbatively, while taking full account of the phase.
For a vanishing modulus Eqs.~(\ref{eq:radial1}) decouple and the radial wave
functions assume the simple form
	\[
		\psi_{\pm}^0(\rho)=A_{\pm}J_{\mu\pm\frac{\nu}{2}}(g
		\sqrt{1\pm \varepsilon}\,\rho),
	\]
where $J$ is the Bessel function. The eigenvalues $\varepsilon$ must be
determined from the boundary condition at the border of a normalization disk of
radius $R$, and they form a continuum for each value of $\mu$ as
$R\rightarrow\infty$. The $\nu\frac{\pi}{2}$ phase shift can be easily seen
from the asymptotic behavior of $J_m(x)$,
	\[
		J_m(x)\sim\sqrt{\frac{2}{\pi x}}\cos\left[{\textstyle
			x-\left(m+\frac{1}{2}\right)\frac{\pi}{2}}\right]\qquad(x\gg m),
	\]
remembering that $\mu=n+\frac{\nu}{2}$ with integer $n$. The first-order
perturbation theory in $\delta$ gives the energies of the core states as
	\[
		\varepsilon_{\mu}=\varepsilon+2\int_0^R d\rho\,\rho\,\delta(\rho)
		\psi_+^0(\rho)\psi_-^0(\rho).
	\]
Because of the phase shift, the pairing matrix element is qualitatively
different for even and odd values of $\nu$. For even $\nu$ and small
$\varepsilon$, $\psi_+^0$ and $\psi_-^0$ are either in phase or out of phase by
$\pi$, and the integrand is thus either positive or negative, leading to a
maximal pairing energy of order $\Delta_{\infty}$. For odd $\nu$, on the
contrary, the integrand oscillates about zero and the resulting pairing energy
is minimal. This mechanism is illustrated in Fig.~\ref{fig4}, using the exact
eigenstates rather than $\psi_{\pm}^0$. The solutions $\psi_{\pm}$ as well as
the integrand of the pairing matrix element are displayed for the lowest
allowed value of $\mu$ and for winding numbers $\nu=-1$ and $-2$. The numerical
values of the kinetic and pairing energies, defined as
	\begin{eqnarray*}
		\varepsilon_{\text{kin}}&=&\langle\psi_+|L_+|\psi_+\rangle-
		\langle\psi_-|L_-|\psi_-\rangle\\
		\varepsilon_{\text{pair}}&=&2\langle\psi_+|\delta|\psi_-\rangle,
	\end{eqnarray*}
are also indicated.

The analysis of the numerical results at all angular momenta shows that the
energy eigenvalues in Figs.~\ref{fig1} and \ref{fig3} are dominated by the
pairing energy. Due to the cancellation of the electron and hole contributions,
the kinetic term remains small and has a weak $\mu$ dependence. The structure
of the eigenvalue spectra, and in particular the occurrence of low-energy
states at high angular momentum for $|\nu|>1$, can thus be qualitatively
understood by considering the evolution of the pairing matrix element with
$\mu$. With increasing $\mu$, an additional phase shift appears between the
radial electron and hole wave functions at short distances (as can be seen from
the functions $\psi_{\pm}^0(\rho)$, the phase shift always tends to
$\nu\frac{\pi}{2}$ asymptotically, but it is a function of $\mu$ for
$\rho<\mu/g$). This has the effect of increasing the matrix element towards its
maximum value of $\Delta_{\infty}$. For the lower branch in the $\nu=-2$
spectrum in Fig.~\ref{fig3}, there is a $\mu>0$ such that the total phase shift
is close to $-\frac{\pi}{2}$, and the matrix element thus nearly vanishes.

In the Appendix I derive an approximate solution for the bound states of the
$s$-wave vortex. To lowest order in $\mu\nu/g\rho_c$ the eigenvalues are found
to be
	\begin{equation}\label{eq:eigenvalues}
		\frac{E_{\mu}}{\Delta_{\infty}}\approx\frac{-\frac{\mu\nu}{g\rho_c}
		+(2m+1+\nu)\frac{\pi}{2}}{1+g\delta\rho_c},
	\end{equation}
where $m$ is any integer such that $|E_{\mu}|<\Delta_{\infty}$ and $\rho_c$ is
some effective core radius in units of $\xi$. For $\nu=-1$,
$g\delta=\frac{2}{\pi}$, and $\rho_c=1$, one recovers the famous Caroli-de
Gennes-Matricon finding,\cite{Caroli-64} namely that for each $\mu$ there is a
unique solution within the gap ($m=0$) with energy
$E_{\mu}\approx\mu\frac{\Delta_{\infty}^2}{E_{\text{F}}}$.

The term $\nu\frac{\pi}{2}$ in Eq.~(\ref{eq:eigenvalues}) can be traced back to
the phase difference discussed above (see the Appendix). Only for odd values of
$\nu$ does the term in parentheses disappear for some $m$, hence a branch of
bound states crosses the Fermi level at $\mu=0$, contributing to the low-energy
LDOS in the core. In this case the gap between the lowest-energy excitations in
the core is $\Delta
E_{\mu=\pm1/2}=\frac{\Delta_{\infty}^2}{E_{\text{F}}}\frac{\pi^2\nu}
{2\rho_c(\pi+2\rho_c)}$, where $\rho_c$ itself is a function of $\nu$. The
$\nu$-dependence of $\Delta E_{\mu=\pm1/2}$ may be estimated by demanding that
the ``hole'' in $\Delta(\vec{r})$ in Eq.~(\ref{eq:profile}) and in the model
calculation of the Appendix have identical volumes, namely $\pi\rho_c^2$. We
then obtain $\rho_c^2=2|\nu| \int_0^{\infty}dx\,x(1-\tanh^{|\nu|}x)
\approx|\nu|^{2\alpha}$ with $\alpha=0.78$. Hence $\Delta E_{\mu=\pm1/2}$ is a
decreasing function of $|\nu|$:
	\begin{equation}\label{eq:gap-odd}
		\Delta E_{\mu=\pm\frac{1}{2}}\approx
		\frac{\Delta_{\infty}^2}{E_{\text{F}}}\,\frac{|\nu|^{1-\alpha}}{2}
		\frac{\pi^2}{\pi+2|\nu|^{\alpha}}\qquad(\nu\text{ odd}).
	\end{equation}
For even values of $\nu$ there is always a large gap at $\mu=0$, which is given
by $\Delta E_{\mu=0} = \Delta_{\infty}\pi/(1+g\delta\rho_c)$ in this model. It
should be noted that the presence (absence) of this large gap for even (odd)
$\nu$ is due to the absence (presence) of the flux-induced topological
frustration, and does not depend on the order-parameter profile; however, the
width of this gap does depend on the order-parameter profile through the
effective core radius $\rho_c$. Using $\rho_c\approx|\nu|^{\alpha}$ and
assuming the BCS relation $g\delta=\frac{2}{\pi}$ holds, we find that $\Delta
E_{\mu=0}$ is also a decreasing function of $|\nu|$,
	\begin{equation}\label{eq:gap-even}
		\Delta E_{\mu=0}\approx\Delta_{\infty}\frac{\pi^2}{\pi+2|\nu|^{\alpha}}
		\qquad(\nu\text{ even}),
	\end{equation}
in good agreement with the numerical results in Fig.~\ref{fig3}.

\section{The isolated $\bm{D}$-wave vortex}

Within the BdG theory, there is an important qualitative difference between the
core spectra of isolated vortices in $d$-wave and $s$-wave superconductors. The
energy spectrum of $s$-wave vortices is discrete (but looks continuous in
experiments because of small interlevel spacing and thermal broadening), while
numerical calculations suggest that the energy spectrum is continuous in the
$d$-wave case,\cite{Franz-98} although there are conflicting opinions in this
respect.\cite{Morita-97} A complete analytical solution would be useful to
address this problem, but to my knowledge none has been published so far. The
main difficulty comes from the nonlocality of the $d$-wave gap operator:
gauge-invariant generalizations of the lattice $d$-wave gap to the continuum
model are rather complicated.\cite{Vafek-01} In this section, the dependence of
the core energy spectrum on the gap profile and vortex winding number is
investigated numerically in the $d$-wave case. The issue of bound versus
extended core states is not directly relevant here.

Instead of a continuum model I consider a two-dimensional nearest-neighbor
lattice model. The vortex order parameter is taken to be
	\[
		\Psi(\vec{r}_i,\vec{r}_j)=\left\{\begin{array}{ll}
		\frac{1}{4}\Delta(|\vec{R}_{ij}|)\cos(2\tau_{ij})e^{i\nu\theta_{ij}}&
		\text{if } |\vec{r}_i-\vec{r}_j|=a\\ [1em]
		0&\text{otherwise,}\end{array}\right.
	\]
where $\vec{r}_i$ denotes a lattice site, $a$ is the lattice parameter,
$\vec{R}_{ij}=\frac{1}{2}(\vec{r}_i+\vec{r}_j) =
|\vec{R}_{ij}|(\cos\theta_{ij},\,\sin\theta_{ij})$,
$\vec{r}_i-\vec{r}_j=a(\cos\tau_{ij},\,\sin\tau_{ij})$, and $\Delta(r)$ is
given by Eq.~(\ref{eq:profile}). The LDOS in the core is calculated using the
Green's function formalism:
	\begin{equation}\label{ldos}
		N(\vec{r}_i,E)=-\frac{2}{\pi}\,\text{Im}\,G_{ii}(E+i\Gamma).
	\end{equation}
The Green's function is given by the Gorkov equations
	\begin{subequations}\label{eq:Gorkov}
	\begin{eqnarray}
		\label{eq:G}
		G^{ }_{ij}(\omega)&=&G^0_{ij}(\omega)+\sum_{kl}G^0_{ik}(\omega)
			\Sigma^{ }_{kl}(\omega)G^{ }_{lj}(\omega)\\
		\label{eq:G0}
		G^0_{ij}(\omega)&=&\frac{1}{N^2}\sum_{\vec{k}}
			\frac{e^{i\vec{k}\cdot(\vec{r}_i-\vec{r}_j)}}
			{\omega-\varepsilon_{\vec{k}}}\\
		\Sigma_{ij}(\omega)&=&\sum_{kl}\Psi(\vec{r}_i,\vec{r}_k)G^0_{lk}(-\omega)
			\Psi^*(\vec{r}_l,\vec{r}_j)
	\end{eqnarray}
	\end{subequations}
with $N^2$ the number of $\vec{k}$ points and $\varepsilon_{\vec{k}}$ the
dispersion. The details of $\varepsilon_{\vec{k}}$ are not important in the
context of this study. However, the presence of a van-Hove singularity in the
gap region provides additional spectral weight, which might be unequally
distributed among the various peaks in the spectra, thus complicating their
interpretation. In order to avoid such difficulties, the simple
nearest-neighbor form $\varepsilon_{\vec{k}}=-2t(\cos k_xa+\cos k_ya)-\mu$ is
used, with the chemical potential set to $\mu=t$ so that the van-Hove
singularity does not influence the spectra in the gap region.
Eq.~(\ref{eq:Gorkov}) is solved by first computing $G^0_{ij}(\omega)$ for a
large system ($N=1024$), taking advantage of the translational invariance in
Eq~(\ref{eq:G0}). Then the inhomogeneous terms $\Sigma_{ij}(\omega)$ and
$G_{ij}(\omega)$ are calculated on a smaller $M\times M$ system ($M=51$) with
the vortex at the center. With this method the finite-size effects do not
contaminate the free propagator, and a good spectral resolution can be
achieved. The hopping integral is set to $t=5\Delta_{\infty}$ (a value typical
for high-$T_c$ materials) and the energy broadening is
$\Gamma=\Delta_{\infty}/50$.

\begin{figure}[t!]
\includegraphics[width=7.5cm]{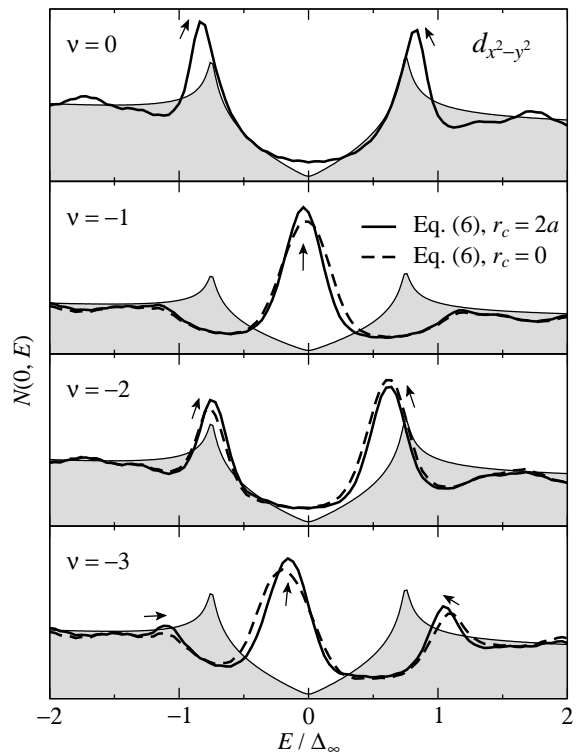}
\caption{\label{fig5}
Local density of states in the core of $d$-wave vortices with increasing
winding numbers. The solid lines correspond to vortices having a normal core
given by Eq.~(\ref{eq:profile}), and the dashed lines correspond to vortices
without normal core. The case $\nu=0$ corresponds to a normal core without
supercurrent. The arrows show the displacement of the peak maxima when the
system size increases from $M=41$ to $M=51$. The bulk DOS is shown in grey, and
provides a common scale to compare the spectra at different $\nu$. Note that
the bulk coherence peaks are not located at $E=\pm\Delta_{\infty}$ due to the
large value of the chemical potential $\mu=-t$.}
\end{figure}

The LDOS calculated at the vortex center with $r_c=2a$ and $r_c=0$ and for
various winding numbers is displayed in Fig.~\ref{fig5}. The LDOS for a
``normal core'' of radius $2a$ without supercurrent is also shown and denoted
as $\nu=0$. One can see several striking similarities with the $s$-wave vortex.
For $|\nu|\geqslant1$, the suppression of the order parameter has a very small
effect on the LDOS in the core. This is confirmed by the case $\nu=0$, which
shows that a local suppression of $\Delta(r)$ is unable to induce low-energy
states. (The small flat DOS at $E=0$ for $\nu=0$ and $\nu=-2$ is probably a
finite-size effect.\cite{size-effet}) For $\nu=0$, there is a transfer of
spectral weight from high to low energy at $E\approx\Delta_{\infty}$, resulting
in two sharp states near the gap edges. Increasing the system size from $M=41$
to $M=51$, these states sharpen while moving to slightly lower binding energy,
as indicated by the small arrows in the figure. At convergence, these states
would presumably lie within the bulk gap, as in the $s$-wave case. For
$|\nu|\geqslant1$, one observes a strong even/odd effect. Vortices with odd
$\nu$ have a zero-bias conductance peak (ZBCP), which is absent in even
vortices. The ZBCP sharpens with increasing system size, but its energy is well
converged at the system size considered. For $|\nu|\geqslant2$, there are
additional states at high energy, which are the analogs of the high-energy
states near $\mu=0$ in the $s$-wave case (see Fig.~\ref{fig3}). One can observe
that the sensitivity of the peaks to system size increases with energy,
suggesting that the states at higher energy are less localized.

\begin{figure}[t!]
\includegraphics[width=8.0cm]{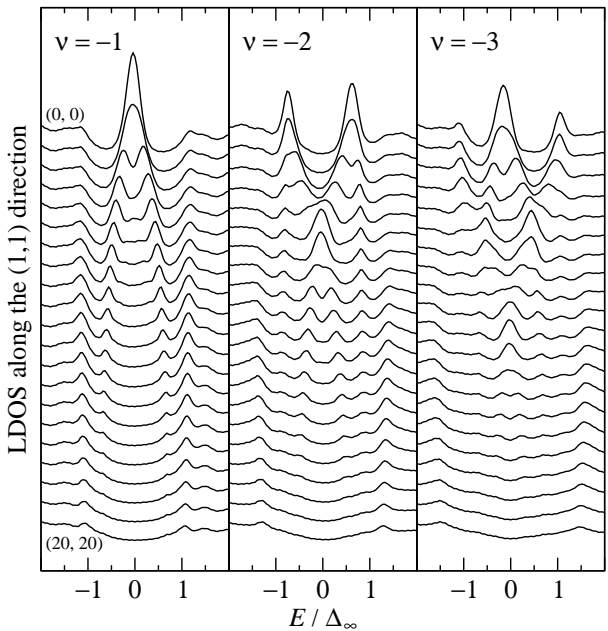}
\caption{\label{fig6}
LDOS along the nodal direction for $d$-wave vortices of increasing winding
numbers. The curves show the LDOS at all sites between $(0,\,0)$ and
$(20,\,20)$, and are offset vertically for clarity.}
\end{figure}

In the neighborhood of the vortex axis, the LDOS has many characteristics in
common with the LDOS of $s$-wave vortices. In Fig.~\ref{fig6}, one can see that
in the case $\nu=-1$ the ZBCP at the core center splits into two symmetric peaks
at larger distances, very much like the high angular momentum electron-hole
excitations in the $s$-wave case (compare Fig.~\ref{fig2}). These excitations
have been the focus of a recent self-consistent calculation.\cite{Zhu-01} In the
doubly-quantized vortex, the two peaks near the gap edges vanish slowly as the
distance from the core increases, while a ZBCP develops, which is maximum at a
distance $r=5\sqrt{2}a$. The latter again decomposes into two symmetric
excitations at larger distances. The resulting LDOS is thus very similar to that
of a doubly-quantized $s$-wave vortex (see Fig.~4 of
Ref.~\onlinecite{Virtanen-99}), and can be readily interpreted on the basis of
Fig.~\ref{fig3}. A similar analysis shows that all of the features of the
$s$-wave vortex with $\nu=-3$ (Fig.~4 of Ref.~\onlinecite{Virtanen-99}) can be
distinguished in the LDOS of the $\nu=-3$ $d$-wave vortex (Fig.~\ref{fig6}). The
LDOS exhibits some anisotropy around the vortex, but the differences between the
nodal and antinodal directions are small. In particular, all the features
discussed in Fig.~\ref{fig6} are also present in the antinodal direction.

\section{Perturbations of the order-parameter phase}
\label{sect:phase}

The results of the previous sections suggest that the core states have a
topological origin. An implication of this is that the states must not be
suppressed by changes in the phase of the order parameter, as long as these
changes preserve the topological defect. Conversely, the core states should be
strongly affected by the interaction with an antivortex, since the latter
annihilates the effect of the topological defect on orbits larger than the
vortex-antivortex distance. In this section, I investigate the effects of some
perturbations of the phase field on the core states in a $d$-wave
superconductor, starting with the perturbation induced by a nearby antivortex.

\begin{figure}[t!]
\includegraphics[width=8.0cm]{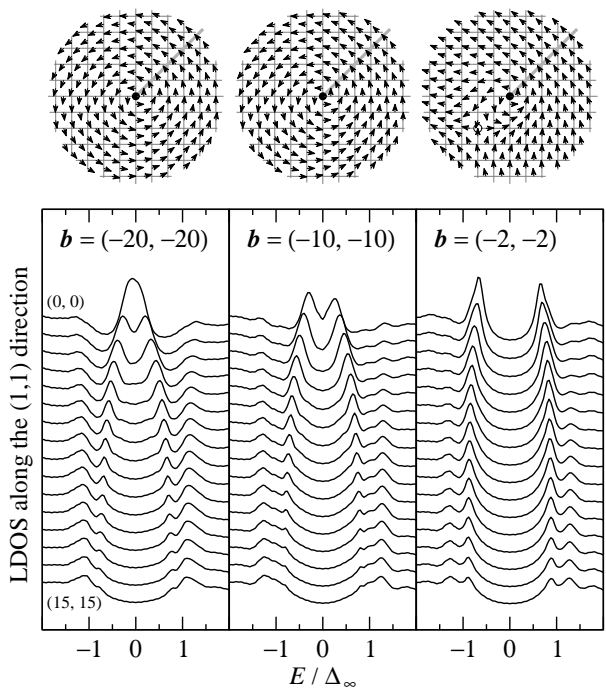}
\caption{\label{fig7}
LDOS along the nodal direction (indicated by a thick gray line in the top part)
for a vortex-antivortex pair in a $d$-wave superconductor for various
vortex-antivortex separations. The antivortex (white dot, only visible at the
shortest distance) is at position $\vec{b}$ with respect to the vortex
(black dot). The order-parameter phase in the region of the vortex center
(excluding the $d_{x^2-y^2}$ symmetry factor for clarity) is represented in the
form of a small arrow on each bond of the square lattice.}
\end{figure}

The order-parameter for a vortex-antivortex pair was taken as the normalized
product of the order parameters associated with each constituent (i.e. two
vortices with winding numbers $1$ and $-1$ and $r_c=2a$). The vortex is located
at the origin and the antivortex at position $\vec{b}=(-b,\,-b)$. In
Fig.~\ref{fig7} is shown the LDOS at the vortex center and along the diagonal,
in the direction opposite to the vortex-antivortex direction. The phase field
around the vortex is also displayed. For $b=20$, the LDOS is similar to the LDOS
of an isolated singly-quantized vortex (Fig.~\ref{fig6}). However, although the
phase field is barely modified by the antivortex in the region where the LDOS is
calculated, the latter is considerably broadened: the height of the ZBCP is only
$71\%$ of the corresponding height in the isolated vortex, while the gap
integrated spectral weight is conserved within $2\%$. Note also that the ZBCP is
already split at site $(1,\,1)$, unlike in the isolated case where this
splitting occurs first at site $(2,\,2)$. Reducing the vortex-antivortex distance
($b=10$), the ZBCP at site $(0,\,0)$ also splits. At the same time, the energy
separation between the two states away from the core center is increased.
Finally, at $b=2$ the ZBCP disappears completely and the LDOS resembles the
LDOS of a $\nu=0$ ``vortex'' (see Figs.~\ref{fig5} and \ref{fig2}).

The spectra in Fig.~\ref{fig7} show that the formation of the core states is a
nonlocal process: the LDOS at a particular site depends on the phase winding in
a region much larger than the core radius $r_c$. In fact the height of the ZBCP
at $(0,\,0)$ reaches $95\%$ of its value in the isolated vortex for
vortex-antivortex separations as large as $\sim70a$ ($b=50$). This value sets
an upper bound for the size of the aforementioned region. On the other hand, a
clear splitting of the ZBCP at site $(0,\,0)$ occurs for $b=19$ or less, which
sets a lower bound to $27a$. These numbers can be compared with the spatial
extension of the core states. In the $s$-wave vortex the latter is typically
$\ell\approx 2E_{\text{F}}/(k_{\text{F}}\Delta_{\infty})$ for a state at $E=0$
[see Eq.~(\ref{eq:decay})]. Using $E_{\text{F}}=5t$ (the position of the Fermi
energy with respect to the bottom of the band in the calculations),
$k_{\text{F}}=\pi/(2a)$, and $t=5\Delta_{\infty}$ one obtains $\ell\approx32a$,
which falls within the bounds deduced from the numerical simulations. Thus,
although the typical localization radius of the bound states might be different
in $d$-wave and $s$-wave superconductors,\cite{Franz-98} one may conclude that
the LDOS in the vortex core ``feels'' the presence of an antivortex (or another
vortex in a vortex lattice) at any distance shorter than the extension of the
bound states.

\begin{figure}[t!]
\includegraphics[width=8.0cm]{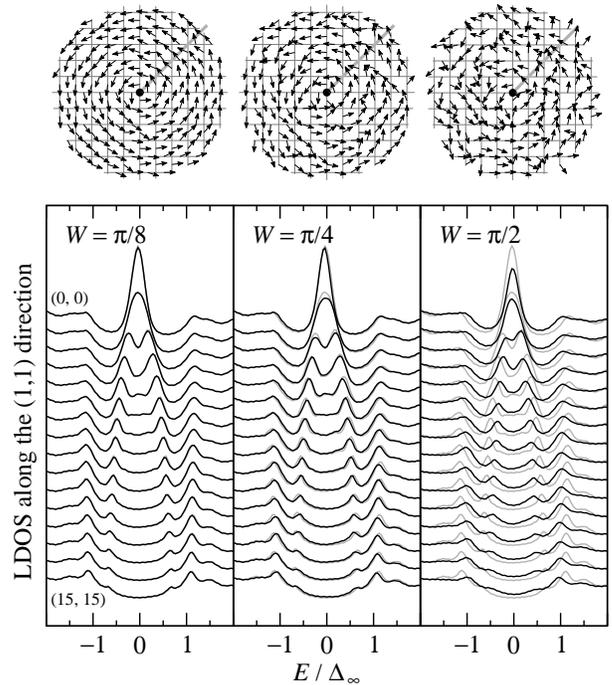}
\caption{\label{fig8}
LDOS along the nodal direction for a $d$-wave singly-quantized vortex subject to
random phase disorder. $W$ characterizes the strength of disorder. The phase
field in the core region is shown in the top part. The spectra corresponding to
the ordered case ($W=0$) are shown in grey for comparison.}
\end{figure}

The sensitivity of the LDOS to perturbations of the phase which preserve the
topological defect was studied by randomly disordering the phase field, i.e.
replacing the order parameter $\Psi$ of the isolated vortex by $\Psi e^{i x}$
where $x$ is a random Gaussian variable centered at $x=0$ with variance $W$
(FWHM). The resulting LDOS is displayed in Fig.~\ref{fig8}. At $W=\pi/8$ and
$W=\pi/4$, the spectra are practically unchanged with respect to the ordered
case. No new structure is induced in the spectra up to the strongest disorder
considered. At $W=\pi/2$, though, the ZBCP is reduced (but {\em not\/}
broadened) and the energy separation between the two electron-hole excitations
away from the core is decreased. This might be attributed to the loss of
rotational symmetry, which leads to a stronger interaction between angular
momentum eigenstates, and to a redistribution of spectral weight. We note that
the integrated LDOS is conserved within $3\%$ at all sites and for all disorder
strengths considered.

\section{Discussion}

The numerical results reported in Figs.~\ref{fig2} and \ref{fig5} imply that
the suppression of the modulus $\Delta(\vec{r})$ of the order parameter in the
core can be safely ignored when discussing the mechanism of formation of the
vortex states in $s$- and $d$-wave superconductors. The main reason is the
small magnitude of the superconducting gap with respect to the Fermi energy: a
local suppression of $\Delta(\vec{r})$ is a {\em very weak\/} perturbation for
excitations at the Fermi surface, which does not provide a substantial gain of
pairing energy for a localized Bogoliubov excitation (unless the localization
radius is smaller than the core radius, which in turn costs a large kinetic
energy). Making use of Eq.~(\ref{eq:eigenvalues}) with $\nu=0$, one can see
that a local suppression of $\Delta(\vec{r})$ can induce low-energy states in
the limit $\Delta_{\infty}/E_{\text{F}}\gg(k_{\text{F}}r_c)^{-1}$, which is
never attained in the BCS superconductors for which
$\Delta_{\infty}/E_{\text{F}} \approx \frac{2}{\pi}(k_{\text{F}}r_c)^{-1}$.

Thus the origin of the bound states must be searched in the phase
$\chi(\vec{r})$ of the order parameter. Far from the vortex, the slow variation
of $\chi(\vec{r})$ is known to induce a small density of states in the gap by
the Doppler-shift effect.\cite{Volovik-93a} The Doppler-shift approximation is
not valid in the core region because there the superfluid velocity is too
large, but also because this approximation neglects the topological defect
associated with the phase winding. A comparison of the right panels in
Figs.~\ref{fig7} and \ref{fig8} unambiguously shows that the microscopic
details of $\chi(\vec{r})$ are much less relevant to the formation of the
vortex LDOS than the ``macroscopic'' phase winding. This is further evidenced
by the qualitative differences between even-$\nu$ and odd-$\nu$ vortices. The
mechanism described in Sect.~\ref{sect:interpretation} provides a natural
explanation to these results, since the interference between the electron and
the hole, which is responsible for the formation of the bound states, is a
global property of the wave function and is not destroyed by local
perturbations of $\chi(\vec{r})$. The issue of self-consistency in the order
parameter (modulus and phase) is one of the major obstacles to a complete
analytical calculation of the vortex-core LDOS. Nevertheless, the results of
the present study suggest that the LDOS calculated from the simplest possible
``trial'' order parameter, i.e. a uniform modulus
$\Delta(r,\,\theta)\equiv\Delta_{\infty}$ and a geometric phase
$\chi(r,\,\theta)=\nu\theta$, should exhibit all the characteristic features of
the exact solution. (An interesting exception might be the case where
subdominant order parameters of different symmetries are brought about by the
self-consistency.)

In the presence of several topological defects, the winding $\nu$ in
Eq.~(\ref{eq:nu}) depends on the number of defects contained in the path of
integration. Based on the analysis in Section~\ref{sect:phase}, I tentatively
argue that the structure of the vortex core LDOS is determined primarily by the
average vorticity in the region occupied by the core states. This statement was
checked by considering a variety of vortex and antivortex configurations. For
example the effect of a $\nu=2$ antivortex on the LDOS in the core of a
$\nu=-1$ vortex was studied. The antivortex was located at position
$\vec{b}=(-b,\,-b)$ as in Fig.~\ref{fig7}. For small ($b/a<2$) and large
($b/a\gtrsim20$) vortex-antivortex separations the LDOS was found to be close
to that of an isolated vortex. Near $b=5a$ the spectrum was similar to the
topmost case in Fig.~\ref{fig5}, corresponding to $\nu=0$. Finally in the other
regions the LDOS exhibited intermediate shapes. In this geometry the vorticity
is $-1$ in the region $r<b\sqrt{2}$ and $+1$ in the region $r>b\sqrt{2}$, so
that the average vorticity in a region of radius $R$ is
$\bar{\nu}=1-4\left(\frac{b}{R}\right)^2$ if $b\sqrt{2}<R$ and $-1$ if
$b\sqrt{2}>R$. Thus $|\bar{\nu}|\approx1$ for small and large $b$, whereas
$|\bar{\nu}|\approx0$ for $b\sim R/2$.

In the normal state of the superconducting cuprates, the behavior of the
high-frequency optical conductivity\cite{Corson-99} and the presence of a large
Nernst signal\cite{Xu-00} have been generally attributed to vortex excitations,
in the form of unbound vortex-antivortex pairs. Recently, Lee\cite{Lee-03}
argued that such vortices have to be ``cheap'', i.e. must be free of core
states in order to have a formation energy comparable to the thermal energy
$\sim k_{\text{B}}T_c$. Experimentally, there is indeed convincing evidence
from STM\cite{Aprile-95,Renner-98,Pan-00,Hoogenboom-01} and
NMR\cite{Mitrovic-01,Kakuyanagi-02} measurements that the core states are
suppressed at $T<T_c$ in the cuprates, pointing to a failure of the simple
$d$-wave BCS theory in the superconducting state. Lee pointed out that a
straightforward extension of the BCS theory which includes phase fluctuations
is unable to produce ``cheap'' vortices in the normal state. This argument
relies on the assumption that the average vortex core LDOS in a
vortex-antivortex soup is similar to the LDOS of an isolated vortex. As
Fig.~\ref{fig7} shows, however, the vortex core states can be very efficiently
suppressed in a ``BCS'' vortex anti-vortex pair, even when the
vortex-antivortex separation is much larger than the core radius. It is likely
that the same phenomenon also occurs in more sophisticated models of the
high-$T_c$ superconductors. A systematic investigation of the core energy as a
function of the vortex density in a state of fluctuating vortex-antivortex
pairs would thus be helpful to elucidate the nature of the normal state in the
cuprates.

In this study, the effect of the order-parameter symmetry on the properties of
the vortex states was left aside, and the emphasis was put on the similarities
of the LDOS in $s$- and $d$-wave vortices, which illustrates the key role of
the vorticity in both cases. It is well known that a subdominant order
parameter can induce qualitative changes in the LDOS of BCS
vortices.\cite{Franz-98} Furthermore, the delicate question of the spatial
extension of the core states in the $d$-wave case (exponentially localized or
extended) was not addressed. These issues are a good motivation to search for a
realistic analytical solution of the BdG equations for vortices in
superconductors with nonlocal pairing.

\section{Conclusion}

In the BCS theory, the superconducting state is characterized by a complex
order parameter, with a modulus $\Delta(\vec{r})$ related to the superfluid
density, and a phase $\chi(\vec{r})$ related to the supercurrent. Although the
electronic states bound to vortices have been generally attributed to the
suppression of the superfluid density in the core region, it was found that the
bound states are completely formed when this suppression is overlooked. In
order to explain the formation of the core states, a new mechanism was
proposed, which relies upon the influence that the vortex supercurrent exerts
on the Bogoliubov excitations. In short, the Bogoliubov excitations localize
onto vortices because the {\em topology\/} of the latter---not the normal
core---gives to the electron and hole components the possibility to avoid one
another and to minimize their pairing energy. The spectral properties of
vortices carrying more than one flux quantum, as well as of vortex-antivortex
pairs, are consistent with this idea. The proposed mechanism of localization
results from the particular form of the coupling mediated by the supercurrent
between the electron and hole components, and is therefore unique to
superconductors.

\begin{acknowledgments}
I am grateful to H. Beck, \O. Fischer, B. Giovannini, S. G. Sharapov, and R. P.
Tiwari for stimulating discussions.
\end{acknowledgments}

\appendix

\section{Approximate solution for the $\bm{S}$-wave vortex}

In this appendix an approximate solution of the BdG equations for an isolated
vortex in a $s$-wave superconductor is derived. This solution differs somewhat
from those given in Refs.~\onlinecite{Caroli-64} and \onlinecite{Bardeen-69},
and applies to vortices of arbitrary winding number $\nu$. The purpose here is
to highlight the role of $\nu$ on the vortex-core spectrum. The radial wave
functions is written as
	\begin{equation}\label{eq:wave}
		\psi_{\pm}(\rho)=\text{Re}\,H_{\mu\pm\frac{\nu}{2}}(g_{\pm}\rho)
		f_{\pm}(\rho),
	\end{equation}
where $g_{\pm}=g\sqrt{1\pm\varepsilon}$ and $H_n(x)=J_n(x)+iY_n(x)$ with $J$ and
$Y$ the Bessel functions of the first and second kind, respectively.
Eq~(\ref{eq:radial}) becomes:
	\begin{equation}\label{eq:radial2}
		\pm\frac{1}{g^2}\left[-\frac{d^2}{d\rho^2}-\left(\frac{1}{\rho}
		+2\frac{H'_{\pm}}{H_{\pm}}\right)\frac{d}{d\rho}\right]
		f_{\pm}+\delta(\rho)\frac{H_{\mp}}{H_{\pm}}f_{\mp}=0.
	\end{equation}
$H_{\mu\pm\frac{\nu}{2}}(g_{\pm}\rho)$ is abbreviated as $H_{\pm}$, and
$H'_{\pm}=dH_{\pm}/d\rho$. At this point I make two simplifications. (i) I
consider a square gap profile: $\delta(\rho)=\delta\,\theta(\rho-\rho_c)$. As
shown in the main text, the vortex-core spectrum is only weakly influenced by
the details of $\delta(\rho)$. (ii) I use the approximations
	\begin{eqnarray*}
		&&\frac{1}{\rho}+2\frac{H'_{\pm}}{H_{\pm}}\approx 2ig_{\pm}\\
		&&\frac{H_{\mp}}{H_{\pm}}\approx\sqrt{\frac{g_{\pm}}{g_{\mp}}}
		e^{\pm i\nu\frac{\pi}{2}}\left(1\mp i\frac{\mu\nu}{\rho_c}\frac{g_++g_-}
		{2g_+g_-}\right)e^{\mp i(g_+-g_-)\rho}.
	\end{eqnarray*}
These expressions were obtained from the asymptotic form of the Bessel
functions,
	\[
		H_n(x)\approx\sqrt{\frac{2}{\pi x}}\,e^{i\left[x+\frac{n^2}{2x}-\left(n+
		\frac{1}{2}\right)\frac{\pi}{2}\right]}\qquad(x>n),
	\]
by expanding to first order in $\mu/g$ and $\nu/g$. The approximation (ii) is
therefore inadequate for $\mu\pm\frac{\nu}{2}>g$. Furthermore, in the second
line, the term proportional to $1/\rho$ was evaluated at $\rho=\rho_c$. This is
justified for the calculation of the eigenvalues, since the latter are
determined by matching the wave-functions at $\rho=\rho_c$. Note also that this
approximation will break down if $\rho_c\ll1$. The factor $e^{\pm
i\nu\frac{\pi}{2}}$ in the expression for $H_{\mp}/H_{\pm}$ is the consequence
of the phase shift discussed in Sect.~\ref{sect:interpretation}. With these
simplifications Eq.~(\ref{eq:radial2}) can be solved exactly. I restrict the
analysis to subgap states with $\varepsilon<\delta$. At $\rho<\rho_c$ the
solution is simply $f^<_{\pm}(\rho)=A_{\pm}$ where $A_{\pm}$ are real
constants. At $\rho>\rho_c$, the solution can be written as
	\begin{equation}\label{eq:decay}
		f^>_{\pm}(\rho)=B_{\pm}e^{i(q-g_{\pm})(\rho-\rho_c)}e^{-k(\rho-\rho_c)}
	\end{equation}
with $q=g(\zeta+\frac{1}{2})^{\frac{1}{2}}$,
$k=g(\zeta-\frac{1}{2})^{\frac{1}{2}}$,
$\zeta=\frac{1}{2}(1+\delta^2+\gamma^2-\varepsilon^2)^{\frac{1}{2}}$, and
$\gamma=\delta\frac{(g_++g_-)\mu\nu}{2g_+g_-\rho_c}$. The $B_{\pm}$ are
complex numbers and they are related by
	\begin{equation}\label{eq:sol1}
		\frac{B_-}{B_+}=\sqrt{\frac{g_-}{g_+}}\,
		\frac{\varepsilon-i\sqrt{4\zeta^2-1}}{\delta-i\gamma}\,
		e^{i[(g_+-g_-)\rho_c-\nu\frac{\pi}{2}]}.
	\end{equation}
Introducing the solutions $f_{\pm}^<$ and $f_{\pm}^>$ into Eq.~(\ref{eq:wave})
and matching the wave-function and its derivative at $\rho=\rho_c$ one obtains
	\begin{equation}\label{eq:sol2}
		\frac{B_s}{A_s}=1-\frac{[kJ_s+(q-g_s)Y_s](Y_s+iJ_s)}
		{(q-g_s)(J_s^2+Y_s^2)-(Y_sJ'_s-J_sY'_s)},
	\end{equation}
with $s=\pm$, $J_{\pm}=J_{\mu\pm\frac{\nu}{2}}(g_{\pm}\rho_c)$, and
$Y_{\pm}=Y_{\mu\pm\frac{\nu}{2}}(g_{\pm}\rho_c)$. Eqs.~(\ref{eq:sol1}) and
(\ref{eq:sol2}) provide the relation between the coefficients $A_+$ and $A_-$, as
well as the eigenvalue equation:
	\begin{equation}\label{eq:tan}
		\tan\left[(g_+-g_-)\rho_c-\nu\frac{\pi}{2}\right]=\frac{-\varepsilon+
		\eta\sqrt{\delta^2+\gamma^2-\varepsilon^2}}{
		\varepsilon\eta+\sqrt{\delta^2+\gamma^2-\varepsilon^2}}
	\end{equation}
where
	\[
		\eta=\frac{\delta+\alpha\gamma}{\gamma-\alpha\delta},\qquad
		\alpha=\tan\arg\frac{B_-/A_-}{B_+/A_+}.
	\]
In Eq.~(\ref{eq:tan}) $g_{\pm}$, $\gamma$, and $\eta$ are all functions of
$\varepsilon$. In order to make Eq.~(\ref{eq:tan}) more tractable, we notice
that $\zeta\approx\frac{1}{2}$ and $g_{\pm}\approx g$ since $\delta$, $\gamma$,
and $\varepsilon$ are small numbers. Taking $\zeta=\frac{1}{2}$ and $g_{\pm}=g$
in Eq.~(\ref{eq:sol2}) we obtain $B_s/A_s=1$ and therefore $\alpha=0$.
Eq.~(\ref{eq:tan}) can then be solved to first order in $\varepsilon$:
	\begin{eqnarray}
		\label{eq:eigenvaluesA}
		\frac{\varepsilon}{\delta}=\frac{E}{\Delta_{\infty}}&\approx&\frac{
		-\tan^{-1}\!\left(\frac{\mu\nu}{g\rho_c}\right)+(2m+1+\nu)\frac{\pi}{2}}
		{\left[1+\left(\frac{\mu\nu}{g\rho_c}\right)^2\right]^{-\frac{1}{2}}
		+g\delta\rho_c}\\ [1em]
		\nonumber
		&\approx&\frac{-\frac{\mu\nu}{g\rho_c}+
		(2m+1+\nu)\frac{\pi}{2}}{1+g\delta\rho_c},
	\end{eqnarray}
where $m$ is any integer such that $|E|<\Delta_{\infty}$ and the second line is
valid to first order in $\mu\nu/g\rho_c$. I have checked numerically that the
eigenvalues resulting from Eqs.~(\ref{eq:tan}) and (\ref{eq:eigenvaluesA}) are
in excellent agreement at all $\mu$. Eq.~(\ref{eq:eigenvaluesA}) correctly 
accounts for the qualitative difference between odd-$\nu$ and even-$\nu$
vortices. Due to the approximations made, however, the validity of
Eqs.~(\ref{eq:tan}) and (\ref{eq:eigenvaluesA}) is limited to the region
$|\mu\nu|<g$, and the zero-energy states at $|\mu|\gtrsim g$ in Fig.~\ref{fig3}
are not reproduced.


\begin{thebibliography}{99}
\bibitem{note-R}
The wave functions were normalized on a disk of radius $R=100\,\xi$.
\end{thebibliography}
\end{document}